\documentclass[aps,preprint,superscriptaddress]{revtex4}



\usepackage{natbib}

\usepackage{graphicx}
\usepackage{dcolumn}

\usepackage{amsmath}

\begin{document}


\title{When do improved covariance matrix estimators enhance portfolio optimization? An empirical comparative study of nine estimators}



\author{Ester Pantaleo}
\affiliation{Dipartimento di Fisica, Universit\`a di Bari, I-70126 Bari, Italy}

\author{Michele Tumminello}
\affiliation{Dipartimento di Fisica e Tecnologie Relative, Universit\`a di Palermo, Viale delle Scienze, I-90128 Palermo, Italy}

\author{Fabrizio Lillo}
\affiliation{Dipartimento di Fisica e Tecnologie Relative, Universit\`a di Palermo, Viale delle Scienze, I-90128 Palermo, Italy}
\affiliation{Santa Fe Institute, 1399 Hyde Park Road, Santa Fe, NM 87501, USA}

\author{Rosario N. Mantegna}
\affiliation{Dipartimento di Fisica e Tecnologie Relative, Universit\`a di Palermo, Viale delle Scienze, I-90128 Palermo, Italy}

\date{\today}
 
\begin{abstract}
The use of improved covariance matrix estimators as an alternative to the sample estimator is considered an important approach for enhancing portfolio optimization.  
Here we empirically compare the performance of $9$ improved covariance estimation procedures by using daily returns of 90 highly capitalized US stocks for the period 1997-2007.
We find that the usefulness of covariance matrix estimators strongly depends on the ratio between estimation period $T$ and number  of stocks $N$, on the presence or absence of short selling, and on the performance metric considered.
When short selling is allowed, several estimation methods achieve a realized risk that is significantly smaller than the one obtained with the sample covariance method. This is particularly true when $T/N$ is close to one. Moreover many estimators reduce the fraction of negative portfolio weights, while little improvement is achieved in the degree of diversification. On the contrary when short selling is not allowed and $T>N$, the considered methods are unable to outperform the sample covariance in terms of realized risk but can give much more diversified portfolios than the one obtained with the sample covariance. When $T<N$ the use of the sample covariance matrix and of the pseudoinverse gives portfolios with very poor performance.
\end{abstract}


\maketitle

\section{Introduction}

Portfolio optimization \cite{Markowitz1952,Markowitz1959,Elton95} is one of the main topics in quantitative finance. Markowitz's solution to the portfolio optimization problem, the mean--variance efficient portfolio, relies upon a series of assumptions and is constructed by using first and second sample moments of financial asset returns. Although analytical and elegant, Markowitz solution to the portfolio optimization problem turns out to be highly sensitive to estimation errors of sample moments. For this reason many moment estimators have been proposed to improve the performance of the portfolio optimization. Furthermore the typical outcome of the Markowitz optimization procedure, especially for large portfolios, is characterized by large negative weights for a certain number of assets of the portfolio \cite{Best1992,Green1992,Jagannathan2003}. Negative portfolio weights require to take a short position (selling an asset without owning it) which is sometimes difficult to implement in practice, or forbidden to some classes of investors. For this reason it is quite widespread to constrain portfolio weights in the optimization procedures. 

In the present study, we focus on the role played in the portfolio selection by estimation errors of the second moments of asset returns, both when taking short selling positions is allowed and when it is forbidden. We can ignore estimation errors of asset returns by restricting our attention to the global minimum variance portfolio, where asset returns are not involved \cite{Ingersoll1987}. It is to notice that this choice is not a limiting one. In fact, the global minimum variance portfolio is typically characterized by an out-of-sample Sharpe ratio (the ratio between the portfolio return and its standard deviation, a key portfolio performance measure) which is as good as that of other efficient portfolios \cite{Jorion1985,Jagannathan2003}. Indeed, there is a consensus on the view that benefits of diversification can be achieved from risk reduction  rather than from return maximization \cite{Jorion1985}.  Furthermore, the determination of expected returns is the role of the economist and of the portfolio manager who are asked to generate or select valuable private information, while estimation of the covariance matrix is the task of the quantitative analyst \cite{Ledoit03b}.

The simplest estimator of the covariance matrix of $N$ asset returns is the  sample covariance estimator, which has $N\times (N+1)/2$  ($\sim N^2/2$ when $N$ is large) distinct elements. For an estimation time horizon of length $T$, the number of available data is $N\times T$.
A very common circumstance in portfolio selection is that the number of assets $N$ is of the same order of magnitude as the estimation  time horizon $T$, for example because non stationarity problems arise for large $T$, or because the portfolio is very large. In this case, the total number of parameters to be estimated is of the same  order of magnitude as the total size of  available data. This unavoidable lack of data records generates large estimation errors in the sample covariance matrix, and thus covariance filtering methods are especially useful, in order to reduce the estimation error. Here we discuss and compare the performance of portfolios obtained by using several estimators of the covariance matrix.  We perform the comparison of portfolio selection methods at different time horizons $T$, and we consider the portfolio optimization problem both with and without including short selling constraints. Specifically, we apply portfolio optimization methods to $90$ highly capitalized stocks traded at the New York Stock Exchange (NYSE) during the time period from January 1997 to December 2005. We find the global minimum variance portfolio both with and without short selling constraints at different time horizons. The investment and estimation horizons are chosen to be identical, and range from one month (approximately $T=20$ trading days)  to two years (approximately $T=480$ trading days). We compare the performance of 10 covariance matrix estimators, namely the sample covariance estimator used in the Markowitz optimization, three estimators based on the spectral properties of the covariance matrix \cite{Metha90,Laloux1999,Plerou1999,Rosenow2002,Potters2005}, three estimators based on hierarchical clustering \cite{Anderberg,Mantegna1999,Tumminello2007,Tola2008,ester}, and three estimators based on shrinking procedures  
\cite{Ledoit03b,Ledoit04,ledoit:CC,Jagannathan2003}.

We find that the effectiveness of the last 9 covariance estimators with respect to the sample estimator in portfolio optimization depends on the presence or absence of short selling, on the performance metric considered, and on the ratio $T/N$. 
Specifically, when short selling is allowed, several covariance estimators are able to give portfolios significantly less risky than the Markowitz portfolio. This is particularly true when $T/N$ is close to one in agreement with previous observations that Markowitz portfolio optimization can be quite problematic and ineffective in the $T/N\approx1$ regime \cite{Pafka2002,Pafka2003,Pafka2005,Kondor2007}. Moreover  for a wide range of  $T/N$,  we verify that portfolios obtained by using the proposed estimation procedures have a lower proportion of negative over positive weights (amount of short selling) \cite{Jagannathan2003} than the Markowitz optimal portfolio, especially when $T/N\approx1$. However the degree of effective diversification of the portfolio is similar for different methods (including Markowitz).

The situation is significantly different when short selling is forbidden. When $T>N$ the realized risk of Markowitz portfolio becomes comparable to that of the other portfolios. In this respect the tested estimators are not able to give  portfolios significantly  less risky than the Markowitz one and all the tested estimators have very similar risk. However the portfolios obtained with these estimators are significantly more diversified than the Markowitz portfolio. 

When $T<N$ the inverse of the sample covariance matrix does not exist because it has zero eigenvalues. It has been proposed to use the pseudoinverse to extend the Markowitz optimization to the case $T<N$. We find that portfolios obtained with the pseudoinverse are more risky and less diversified than the other portfolios.    

By comparing portfolios with and without short selling we also verify and generalize the observation that including constraints (such as the no short selling constraint) in the portfolio optimization procedure is similar to perform an unconstrained optimization with a filtered covariance matrix (see Ref. \cite{Jagannathan2003} for shrinkage estimators and Ref. \cite{SchŠfer2010} for some covariance estimators based on spectral properties).

The paper is organized as follows. In Section II we discuss basic aspects of the Markowitz portfolio optimization procedure and set the notation. In Section III we describe the investigated covariance matrix estimators. Section IV presents the data set,  the methodologies used to compare the different portfolios, and the empirical results. Section V concludes.

\section{Markowitz portfolio optimization}
In this section we briefly discuss some basic aspects of portfolio optimization in Markowitz framework. This is also useful to set the notation and state the assumptions made and the methods used.

Given $N$ stocks, at time $t_0$ an investor selects his/her portfolio of stocks by choosing a fraction of wealth $w_i$ to invest in stock $i$, with $i=1,...,N$, in order to have maximum profit and minimum risk from his/her investment at a fixed time $t_0+T$  in the future. The $N$--dimensional column vector of the weights $\bf{w}$ is normalized as ${\bf w}^\top{ \bf 1}_N=1$, where ${\bf 1}_N$ is the $N$--dimensional column vector of ones.
The average return and the variance of the portfolio are 
\begin{align}
r_p=  \bf{w}^\top \bf{m} && \mbox{and} && \sigma_p^2=\bf{w}^\top \bf{\Sigma}\bf{w},
\end{align}
respectively, where ${\bf{m}}$ and ${\bf \Sigma}$ are the $N$--dimensional column vector of mean returns and the
 $N\times N$ covariance matrix  of the stocks, respectively.
Markowitz optimization problem consists in finding the vector $\bf{{w}}$ which minimizes $\sigma_p$ for a given value of $r_p$. The choice of using the standard deviation as a measure of risk is based on the assumption that returns follow a Gaussian distribution. If one does not set any constraint on the value of the weights, allowing them to be either positive or negative, Markowitz solution to the optimization problem \cite{Markowitz1959} is 
\begin{equation}
{\bf w^*}=\lambda {\bf \Sigma}^{-1}{\bf 1}_N+\gamma{\bf \Sigma}^{-1}{\bf m}
\label{markowitz}
\end{equation}
where
\begin{eqnarray}
\lambda=\frac{C-r_p B}{\Delta}~~~~~~~~~\gamma=\frac{r_pA-B}{\Delta}\nonumber \\
A={\bf 1}_N^T{\bf \Sigma}^{-1}{\bf 1}_N~~~~~~B={\bf 1}_N^T{\bf \Sigma}^{-1}{\bf m}\nonumber\\
C={\bf m}^T{\bf\Sigma}^{-1}{\bf m}~~~~~~\Delta=AC-B^2.\nonumber
\end{eqnarray} 
The inverse of the parameter $\gamma$ is usually referred to as risk aversion. 

When $\gamma=0$ (infinite risk aversion), the optimal portfolio is the global minimum variance portfolio and it does not depend on expected returns. Since in this paper we aim to investigate the role of estimation risk of the covariance matrix,  we  focus on the global minimum variance portfolio, as done in Ref.s \cite{Jorion1985,Jagannathan2003,Ledoit03b}, which obviously does not depend on the estimation error of mean returns. Markowitz optimization typically gives both positive and negative portfolio weights  and, especially for large portfolios, it usually gives large negative weights for a certain number of assets \cite{Best1992,Green1992,Jagannathan2003}. A negative weight corresponds to a short selling position (selling an asset without owning it) and it is sometimes difficult to implement in practice or forbidden. For this reason it is common practice to impose constraints to the portfolio weights in the optimization procedure. When one adds constraints on the range of variation of the $w_i$s the optimization problem cannot be solved analytically, and quadratic programming must be used. Quadratic programming algorithms are implemented in most numerical programs, such as {\tt Matlab} or {\tt R}. In the following we will consider the portfolio optimization problem both  with and without the no short selling constraint $w_i\ge 0$  $\forall i=1,\dots,N$. 
 
\section{Covariance matrix estimators}

One of the main problems of portfolio optimization is the estimation of the mean returns vector $\bf{m}$  and covariance matrix ${\bf \Sigma}$. For the  global minimum variance portfolio the investor needs only to estimate ${\bf \Sigma}$. In what follows we estimate the covariance matrix by using past return data. Specifically, at time $t_0$ we estimate the sample covariance matrix of daily returns in the $T$ trading days preceding $t_0$. We then apply the different estimators  and calculate the optimal portfolio. This portfolio is held until time $t_0+T$ when we evaluate its performance. Note that our estimation and investment time horizons are chosen to be the same. We consider three classes of estimators: i) spectral estimators, ii) hierarchical clustering estimators, and iii) shrinkage estimators. 

\subsection{Markowitz direct optimization}

Let us first point out some aspects associated with the Markowitz direct optimization. In this case, the estimator of the covariance matrix at time $t_0$ is the sample covariance matrix estimated on the preceding  $T$ days.  The input to the global minimum variance optimization problem is the inverse of the sample covariance matrix. 
When $T<N$ the inverse of the sample covariance matrix does not exist because of the presence of null eigenvalues. As suggested in the literature (for example in Ref. \cite{Ledoit03b}) in the optimization problem we use the pseudoinverse, also
called generalized inverse \cite{Mardia}, of the covariance matrix. 
Replacing the inverse of the covariance matrix with the pseudoinverse in the optimization problem allows one to get a unique combination of portfolio weights.  It should be noted that, when $T<N$, the optimization problem remains undetermined and the pseudoinverse solution is just a natural choice among the infinite undetermined solutions to the portfolio optimization problem.

In the same regime $T<N$, this problem  does not arise for the other covariance estimators, because they typically give positive definite covariance matrices for any value of $T/N$ including $T/N<1$.  

\subsection{Spectral estimators}

The first class of methods includes three different estimators of the
covariance matrix, which make use of the spectral properties of the correlation matrix. The fundamental idea behind these methods is that
the eigenvalues of the sample covariance matrix carry different
economic information depending on their value.

The first method we consider is the single index model (see for
instance Ref. \cite{Campbell1997,Ledoit03b,ledoit:CC}). In this model stock returns $r_i(t)$ are described by the set of linear equations $r_i(t)=\beta_i f(t)+\varepsilon_i(t)$, $i=1,...,N$ where returns are therefore given by the linear combination of a single random variable,
the index $f(t)$, and of an idiosyncratic stochastic term $\varepsilon_i(t)$. The parameters $\beta_i$ can be
estimated by linear regression of stock return time series on the index return. The
covariance matrix associated with the model is
${\bf S}^{(SI)}=\sigma_{00}{\boldsymbol \beta}{\boldsymbol \beta}^\top+{\bf D}$,
where $\sigma_{00}$ is the variance of the index, ${\boldsymbol \beta}$ is the
vector of parameters $\beta_i$, and ${\bf D}$ is the diagonal matrix
of variances of $\varepsilon_i$. We indicate this method hereafter as SI. It can be shown that this method gives an estimated covariance 
matrix very similar to the one obtained with the method RMT-0 (see below) when only the largest eigenvalue of the sample covariance is assumed to carry reliable economic information.

The other two spectral methods make use of the Random Matrix Theory
(RMT) \cite{Metha90, Laloux1999,Plerou1999}.  Specifically, if the $N$ variables of
the system are i.i.d. with finite variance $\sigma^2$, then in the limit
$T,N \to \infty$, with a fixed ratio $ T/N $, the
eigenvalues of the sample covariance matrix are bounded
from above by the value 
\begin{equation}
\label{rmt}
\lambda_{max}=\sigma^2 (1+N/T+2\sqrt{N/T}),
\end{equation}
where $\sigma^2=1$ for correlation matrices. In most practical cases, one finds that the largest eigenvalue
$\lambda_1$ of the sample correlation matrix of stocks is definitely
inconsistent with RMT, i.e. $\lambda_1\gg \lambda_{max}$. In fact the largest eigenvectors is typically identified with the market mode. To cope with this evidence, Laloux et al. \cite{Laloux1999} propose to modify the null
hypothesis of RMT so that system correlations can be described in terms of a
one factor model instead of a pure random model. Under such a less restrictive null hypothesis the value of
$\lambda_{max}$ is still given by Eq. (\ref{rmt}), but now
$\sigma^2=1-\lambda_1/N$.  Here we consider two different procedures that apply RMT to the covariance estimation problem.

The first procedure has been proposed by Rosenow et al. in Ref. \cite{Rosenow2002} and works as follows. One
diagonalizes the sample correlation matrix and replaces all
the eigenvalues smaller than $\lambda_{max}$ with 0. One then
transforms back the modified diagonal matrix in the standard basis
obtaining the matrix ${\bf H}^{(RMT-0)}$.
The filtered correlation matrix ${\bf C}^{(RMT-0)}$ is obtained by
simply forcing to 1 the diagonal elements of ${\bf H}^{(RMT-0)}$.
Finally the filtered covariance matrix
${\bf S}^{(RMT-0)}$ is the matrix of elements
$\sigma_{ij}^{(RMT-0)}=c_{ij}^{(RMT-0)} \sqrt{\sigma_{ii}\sigma_{jj}}$, where
$c_{ij}^{(RMT-0)}$ are the entries of ${\bf C}^{(RMT-0)}$ and
$\sigma_{ii}$ and $\sigma_{jj}$  are the sample variances of
variables $i$ and $j$, respectively. In the following we will refer to this method as the RMT-0 method.

The second way to reduce the impact of eigenvalues smaller than
$\lambda_{max}$ onto the estimate of portfolio weights has been
proposed  by Potters et al. in Ref.
\cite{Potters2005}. In this
case one diagonalizes the sample correlation matrix and replaces all the
eigenvalues smaller than $\lambda_{max}$  with
their average value. Then one transforms back the modified diagonal
matrix in the original basis obtaining the matrix ${\bf H}^{(RMT-M)}$ of
elements $h_{ij}^{(RMT-M)}$. It is to notice that replacing the
eigenvalues smaller than $\lambda_{max}$ with their average value
preserves the trace of the matrix. Finally, the filtered correlation
matrix ${\bf C}^{(RMT-M)}$ is the matrix of elements
$c_{ij}^{(RMT-M)}=h_{ij}^{(RMT-M)}/\sqrt{h_{ii}^{(RMT-M)}\,h_{jj}^{(RMT-M)}}$.
The covariance matrix ${\bf S}^{(RMT-M)}$ to be used in the
portfolio optimization is the matrix of elements
$\sigma_{ij}^{(RMT-M)}=c_{ij}^{(RMT-M)} \sqrt{\sigma_{ii} \sigma_{jj}}$, where
$\sigma_{ii}$ and $\sigma_{jj}$ are again the sample variances of
variables $i$ and $j$, respectively. We will refer to
this method as the RMT-M method.

\subsection{Agglomerative hierarchical clustering estimators}

The second class of methods comprises three different estimators of the
covariance matrix based on agglomerative hierarchical clustering
\cite{Anderberg}. Agglomerative hierarchical clustering methods are
clustering procedures based on pair grouping where elements are
iteratively merged together in clusters of increasing size according to their degree of similarity.  Hierarchical clustering
procedures therefore depends on the chosen similarity measure between elements of the system. In the present study we consider the correlation as a measure
of similarity between two elements in the system. Hierarchical clustering algorithms work as follows. Given a data set of $N$ time series, at the the beginning each element defines a cluster. The similarity between  two clusters is defined as the
correlation coefficient between the corresponding two time series. Then the two clusters with the largest correlation are merged together in a single cluster. At the second iteration one has to tackle the subtler problem
of defining a similarity between clusters. Different similarities between clusters can be defined, each one characterizing a specific hierarchical clustering procedure. Once the similarity between two clusters is consistently defined, then the two clusters with the largest similarity are merged together, and the procedure is iterated until, after $N-1$ iterations, all the elements are grouped together in one cluster, corresponding to the whole data set.

We consider here three
hierarchical clustering procedures that differ in the definition of similarity between clusters.  In the unweighted pair group method with
arithmetic mean (UPGMA) if a new cluster $L$ is formed from clusters $A$ and $B$, then the similarity between cluster $L$ and any other cluster $F$ is given by
\begin{equation}
\label{upgma}
\rho_{L,F}=\frac{N_A\rho_{A,F}+N_B \rho_{B,F}}{N_A+N_B},
\end{equation}
where $N_A$ and $N_B$ are the number of elements in cluster A and B, respectively.  Within this rule the similarity between cluster L and cluster F
is given by the arithmetic mean of the set $\{\rho_{ij},\, \forall i\in L,\,
\text{and}\, \forall j\in F \}$. In the weighted pair group method
with arithmetic mean (WPGMA) the average is weighted in such a way to
get rid of the possibly different sizes of A and B
\begin{equation}
\label{wpgma}
\rho_{L,F}=\frac{\rho_{A,F}+\rho_{B,F}}{2}.
\end{equation}
Finally, in the Hausdorff linkage cluster analysis 
\cite{ester}, the similarity between cluster $L$ and cluster F is obtained in terms of the Hausdorff distance between
the two clusters
\begin{equation}
\label{Hausdorff}
\rho_{L,F}=\min\{\min_{i\in L}\max_{j\in F}\,\rho_{ij},\max_{i\in
L}\min_{j\in F}\,\rho_{ij}\}.
\end{equation}
The output of any hierarchical clustering procedure is a dendrogram where each node $\alpha_k$ is associated with the similarity $\rho_{\alpha_k}$ between  the two clusters of elements merging together in the node $\alpha_k$. One can
therefore construct a filtered  similarity matrix ${\bf C^<}$
associated with a specific dendrogram as follows. Each entry $\rho^<_{ij}$ of ${\bf C^<}$ is set to $\rho_{\alpha_k}$, where $\alpha_k$ is the node of the dendrogram corresponding to the smallest cluster in which
the elements $i$ and $j$ merge together. The matrix ${\bf C^<}$ is positive definite provided that
its entries are non negative numbers \cite{Tumminello2007} and that the dendrogram does not show reversals
\cite{Anderberg}. The first condition is typically observed in the financial case, while the latter condition is always satisfied by the
UPGMA and the WPGMA, while it could be violated in the Hausdorff method. When
reversals are present in the dendrogram associated with Hausdorff method, we
remove such reversals by using the minimum spanning tree associated
with the hierarchical clustering procedure \cite{TumminelloIJBC2007}.
Since our procedure generates positive definite matrices, they can be interpreted as correlation matrices.
Once ${\bf C^<}$ is constructed, we obtain an estimate of the
covariance matrix by multiplying the entries of ${\bf C^<}$ by the
sample standard deviations. Hierarchical clustering procedures have
been shown to be effective in extracting financial information from
the correlation matrix of stock returns since Ref. \cite{Mantegna1999}. It
is finally to notice that hierarchical clustering methods have already
been considered in portfolio optimization in Ref. \cite{Tola2008}.\\
\subsection{Shrinkage estimators}
The last class of estimators comprises linear shrinkage methods. Linear shrinkage is a well--established technique in high--dimensional inference problems, when the size of data is small compared to the number of unknown parameters in the model. In such cases, the sample covariance matrix is the best estimator in terms of actual fit to the data but it is suboptimal because the number of parameters to be fitted is larger than the amount of data available \cite{Stein}. The idea is to construct a more
robust estimate ${\bf Q}$ of the covariance matrix by shrinking the
sample covariance matrix $\bf{S}$ to a  target matrix
${\bf T}$, which is typically positive definite and has a lower variance. The shrinking is obtained by computing
\begin{equation}
\label{shrinkage}
{\bf Q}=\alpha {\bf T}+(1-\alpha) {\bf S},
\end{equation}
where $\alpha$ is a parameter named shrinkage intensity. We consider three different shrinkage estimates of the covariance matrix, each one characterized by a specific target
matrix. 

The shrinkage to single index uses the target matrix
${\bf T}={\bf S}^{(SI)}=\sigma_{00}{\boldsymbol \beta}{\boldsymbol \beta}^\top+{\bf
D}$, i.e., the single index covariance matrix previously discussed. This target was first proposed in the context of portfolio optimization by Ledoit et al. \cite{Ledoit03b}.
The second method is called shrinkage to common covariance. The target ${\bf T}$ is a
matrix where the diagonal elements are all equal to the average of
sample variances, while non diagonal elements are equal to the average
of sample covariances. In the shrinkage to common covariance the
heterogeneity of stock variances and of stock covariances is therefore minimized. The method has been proposed for the analysis of bioinformatic data in Ref. \cite{Schafer} and, to the best of our knowledge, it has never been used in the context of financial data analysis. 
The third method, termed shrinkage to constant correlation has a more structured target and was used in Ref. \cite{ledoit:CC}.
The estimator is obtained by first shrinking the correlation matrix to a target
named constant correlation, and then by multiplying the shrunk correlation matrix by the sample
standard deviations. The constant correlation target is a matrix with
diagonal elements  equal to one, and off-diagonal elements equal to the average sample correlation between the elements of the
system. As $\alpha$ (the shrinkage intensity) we use the unbiased estimate analytically  calculated in \cite{Schafer}.

\bigskip

In conclusion we consider $10$ covariance
matrix estimators that we label: Markowitz, SI, RMT-0, RMT-M, UPGMA,
WPGMA, Hausdorff, shrinkage to SI, shrinkage to common covariance, and shrinkage to constant
correlation.

\section{Optimization process: empirical results}

In this Section we present repeated portfolio optimizations performed by using the covariance estimators discussed in the previous Section. A set of highly liquid stocks traded at the NYSE is used.

\subsection{Data}
Our dataset consists of the daily returns of $N= 90$ highly capitalized stocks traded at NYSE and included in the NYSE US 100 Index. For these stocks the closure prices are available in the eleven year period from 1 January 1997 to 31 December 2007 \footnote{The data, already preprocessed, were downloaded from {\tt Yahoo Finance}.}. The ticker symbols of the investigated stocks are AA, ABT, AIG, ALL, APA,  AXP,  BA,  BAC, BAX, BEN, BK, BMY, BNI, BRK-B, BUD, C, CAT, CCL, CL, COP, CVS, CVX, D, DD, DE, DIS, DNA, DOW, DVN, EMC, EMR, EXC, FCX, FDX, FNM, GD, GE,  GLW, HAL, HD, HIG, HON, HPQ, IBM, ITW, JNJ, JPM, KMB, KO, LEH, LLY, LMT, LOW, MCD, MDT, MER, MMM, MO, MOT, MRK, MRO, MS, NWS-A, OXY, PCU, PEP, PFE, PG, RIG, S, SGP, SLB, SO, T, TGT, TRV, TWX, TXN, UNH, UNP, USB, UTX, VLO, VZ, WAG, WB, WFC, WMT, WYE, XOM.  As reference index in the SI model and in the shrinkage to single index we use the  Standard \& Poor's 500 index, which is a widely used broadly--based market index.

At time $t_0$ the portfolio is selected by choosing the optimal weights that solve the global minimum variance problem with or without short selling constraints. The input to the optimization problem is the covariance matrix estimator ${\boldsymbol{S}}^{(f)}$ calculated using the $T$ days preceding $t_0$ and obtained with one of the methods (i.e.  $f\in\{$ Markowitz (M), SI , RMT-0, RMT-M, UPGMA, WPGMA, Hausdorff, shrinkage to SI, shrinkage to common covariance, shrinkage to constant correlation$\}$. We call ${\boldsymbol{S}}^{(f)}$ the estimated covariance matrix. For instance, in this notation, $\boldsymbol{S}^{(M)}$ is the sample covariance matrix,  i.e. the one used in Markowitz portfolio optimization. The output of the optimization problem is 
\begin{equation}
{\bf{w}}^{(f)}=\arg\min_{\bf{w}} {\bf{w}}^\top {\bf{S}} ^{(f)} {\bf{w}},
\end{equation}
with the appropriate constraints. The ex post covariance matrix $\hat{\boldsymbol{S}}$ is defined as the sample covariance matrix calculated using the $T$ days following $t_0$.
The predicted portfolio risk is 

\begin{equation}
s_p ^{(f)}=\sqrt{{\bf w}^{(f)\top} {\bf{S}}^{(f)} {\bf w}^{(f)}},
\end{equation}
and the realized portfolio risk is
\begin{equation}
{\hat{s}_p} ^{(f)}=\sqrt{{\bf w}^{(f)\top} {\bf{\hat S}} {\bf w}^{(f)}}.
 \label{post}
 \end{equation}
Thus both $s_p ^{(f)}$ and ${\hat{s}_p} ^{(f)}$ are estimated by using a time window of length $T$.
 The time window $T$  is varied on a wide range.  In our empirical study, we use seven different time windows $T$ of 1, 2, 3, 6, 9, 12, and 24 months. In other words, we select the portfolio monthly ($T\simeq20$), bimonthly ($T\simeq 40$), quarterly ($T\simeq 60$), six-month ($T\simeq 125$), nine-month ($T\simeq 187$), yearly ($T\simeq 250$), and biannually  ($T\simeq 500$).  Since the total number of trading days is 2761, we consider $131$, $65$, $43$, $13$, $21$, $10$, and $8$ portfolio optimizations for the time horizon $T$ equal to $1$, $2$, $3$, $6$, $9$, $12$, and $24$ months, respectively (for the 24 months case, in order to improve the statistics, we repeated the optimization process starting from 1 January 1998). In order to compare risk levels at different time  horizons, we report annualized risks in all figures and tables.

\subsection{Performance estimators}

To evaluate the performance of different covariance estimators we compare portfolio realized risk, portfolio reliability (i.e. the agreement between realized and predicted risk), and effective portfolio diversification of the portfolios ${\bf w}^{(f)}$. From now on we will drop the superscripts $(f)$. Clearly a portfolio is less risky than another when its realized risk is smaller. Therefore our first performance metric is the realized risk. 
Moreover it is important that the portfolio is reliable, i.e., the ex-ante prediction is close to the ex-post observation of the portfolio risk. We consider both an absolute measure, $|\hat{s}_p-{s}_p|$ and a relative, $|\hat{s}_p-s_p|/\hat{s}_p$, measure of reliability. Note that in the relative measure we normalize with respect to the realized risk instead of the predicted risk because the predicted risk can be very small or even zero  when $T<N$. A third aspect to evaluate the performance of a portfolio is a high level of diversification across stocks of the portfolio. Thus we measure the effective portfolio diversification of the different covariance estimator methods. Following \cite{BPbook} the effective number $N_{eff}$ of stocks with a significant amount of money invested 
in is defined as
\begin{equation}
N_{eff}=\frac{1}{\sum\limits_{i=1}^N {\rm w}_i^2}.
\label{partratio}
\end{equation}
This quantity is 1 when all the wealth is invested in one stock, whereas it is $N$ when the wealth 
is equally divided among the $N$ stocks, i.e., ${w}_i = 1/N$.   When all weights are positive, i.e. when short selling is not allowed, the quantity $N_{eff}$ has a clear meaning. On the other hand, when short selling is allowed there might be some ambiguity in the interpretation of $N_{eff}$ \footnote{For instance, consider a portfolio of $N=2M+1$ stocks where $M$ weights are equal to $-x$, $M$ weights are equal to $x$ and the remaining one is equal to $1$ with $x>1$. The weights are normalized to one. In this limit example, the quantity in Eq. (\ref{partratio}) is equal to $N_{eff}=1/(2Mx^2+1)$ which can be much smaller than 1, even if the portfolio is concentrated in $2M$ stocks. This example shows that  $N_{eff}$ is a meaningful measure of portfolio diversification only when short selling is not allowed.}.
For this reason, we introduce another measure of portfolio diversification. Specifically we consider the absolute value of the weights and we compute the smallest number of stocks for which the sum of absolute weights is larger than a given percentage $q$ of the sum of the absolute value of all the weights. In other words we define
\begin{equation}
N_q=\arg\min_{l}  \sum_{i=1}^l |{w}_i|\ge q\sum_{i=1}^N |{ w}_i|.
\end{equation}
In the following we consider $q=0.9$ and we term this indicator as $N_{90}$. $N_{90}$ is the minimum number of stocks in the portfolio such that their absolute weight cumulate to $90\%$ of the total of asset absolute weights.

\subsection{Realized risk and reliability of different covariance estimators}

\begin{figure}[ptb]
\begin{center}
\includegraphics[scale=0.3]{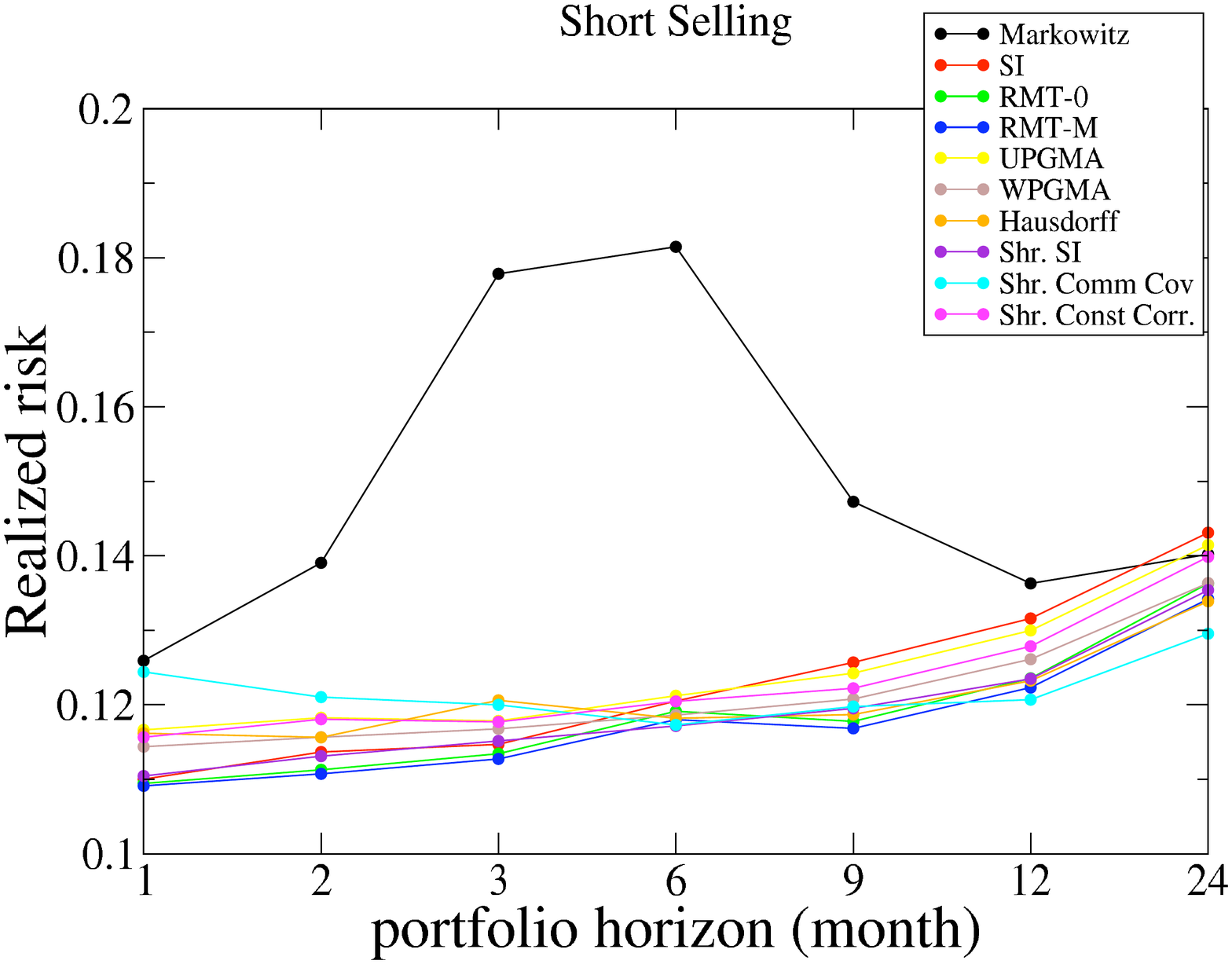}
\includegraphics[scale=0.3]{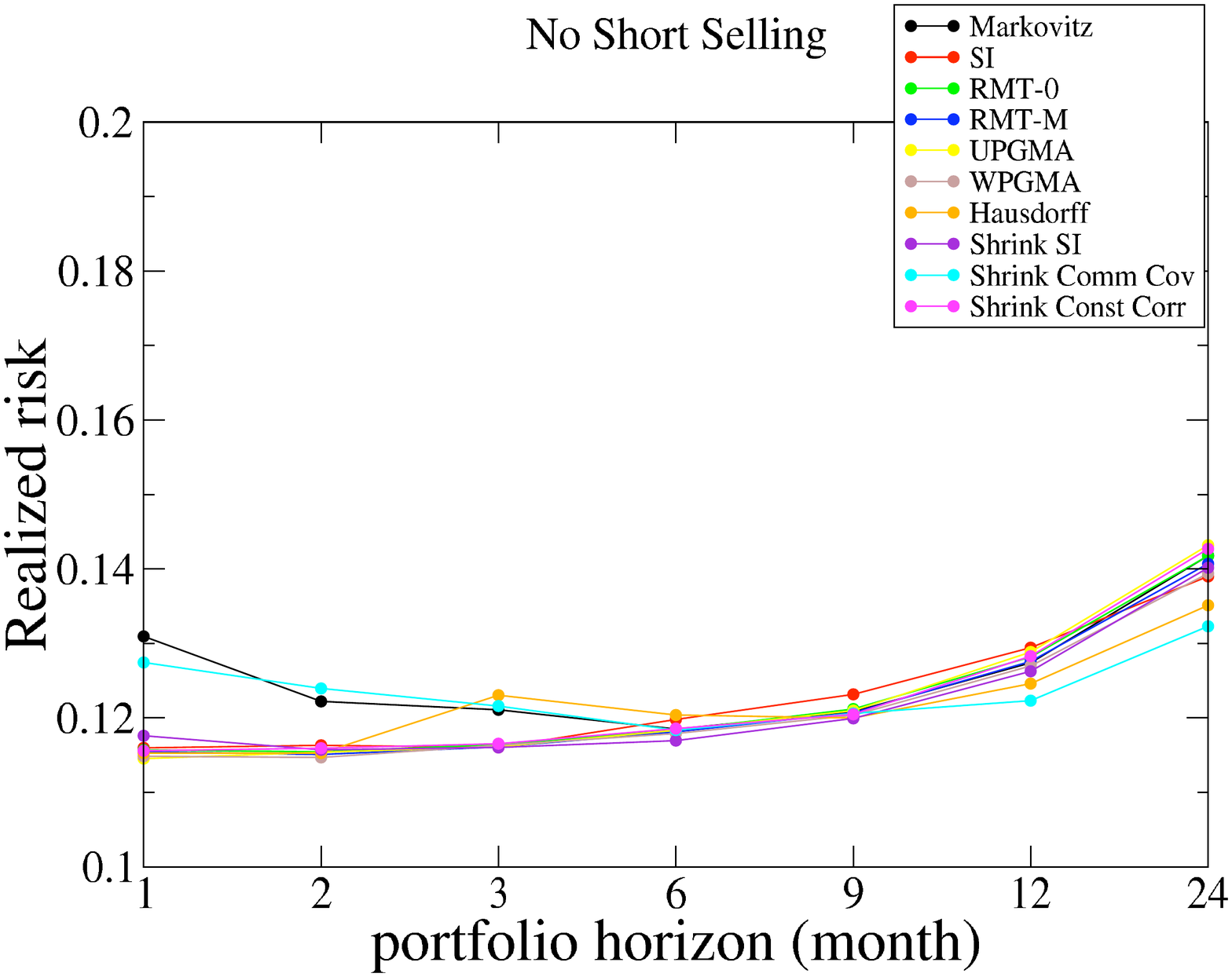}
\end{center}
\caption{Mean realized (annualized) risk $\hat{s}_p$ for  portfolios obtained with the 10 different methods as a function of the horizon $T$. T=1,2,3,6,9,12,24 months correspond to $T/N\approx 0.2,0.4,0.7,2.1,2.8,5.6$, respectively. The top panel considers portfolios where short selling is allowed and the bottom panel considers portfolios where short selling is forbidden.}
\label{realized}
\end{figure} 
In this Section we present the results obtained in repeated portfolio optimization performed by using the covariance estimators described in Section III. Let us first discuss the general qualitative behavior of the realized risk for different estimators, different time horizons $T$ (and thus different ratios $T/N$) and different short selling conditions. Later we perform more rigorous statistical tests.

Figure \ref{realized} shows the mean value of the realized risk (averaged over different portfolio selection times $t_0$) as a function of the time horizon $T$ in the case of short selling (top panel) and no short selling (bottom panel). When short selling is allowed (top panel), the performance of the Markowitz portfolio is very poor and clearly different from that of the portfolios obtained with the other investigated covariance estimators. Markowitz direct optimization procedure gives the highest realized risk at each time window $T$, with the exception of $T=2$ years.
Furthermore, while the realized risk curves of the other optimization procedures are approximately increasing functions of $T$ (except shrinkage to common covariance), the realized risk of the Markowitz portfolio is non monotonic:  the realized risk is very high at $T=3$ and 6 and decreases around those values. The non monotonic behavior of the Markowitz direct optimization method can be explained as follows. When short selling is allowed, a high realized risk at $T\approx 4.5$ months  is expected because  $T \approx N$ (i.e., $T\approx 90$ days=$4.5$ months in our case) is the crossing point from non singular to singular covariance matrices. In fact, in References \cite{Pafka2002,Pafka2003,Pafka2005,Kondor2007}, a divergence of  the realized risk is shown to occur in the limit $T\to\infty$, $N\to\infty$ and $T/N\to 1$ from the right. Here we verify this behavior and we observe the divergence also when $T/N\to 1$ from the left.
From the top panel of Fig. \ref{realized} we can also see how spectral and hierarchical clustering methods show a similar performance in terms of realized risk.  
Shrinkage methods have a performance similar  to that  of the other algorithms, but the shrinkage to common covariance method shows a relatively poorer performance for low values of $T$ while it shows one of the best performances for high values of $T$.

The bottom panel of figure \ref{realized} shows the mean realized risk as a function of the time horizon $T$ when the no short selling condition is imposed. In this case too, the realized risk of all portfolios approximately increases with $T$ except again for the Markowitz optimization and the shrinkage to common covariance method.  Moreover, for $T$ larger than $N$ all the methods are roughly equivalent in terms of realized risk. For $T<N$, Markowitz and shrinkage to common covariance have clearly a high realized risk, while the other methods are again essentially equivalent (with the possible exception of Hausdorff estimator for $T=3$ months).  
Finally, overall, except for the Markowitz portfolio, a comparison of the top and bottom panels of Fig. \ref{realized} shows  that the realized risk of all portfolios turns out to be approximately the same both when constraints on short selling are applied and when they are not.

In the previous analysis we have considered the average realized risk  over repeated optimizations for different time horizons  $T$. Now, we fix $T$ and consider the realized risk time series to explore the role and nature its fluctuations in different market conditions. We compare these time series for different values of the time horizon $T$.  
\begin{figure}[ptb]
\begin{center}
\includegraphics[scale=0.5]{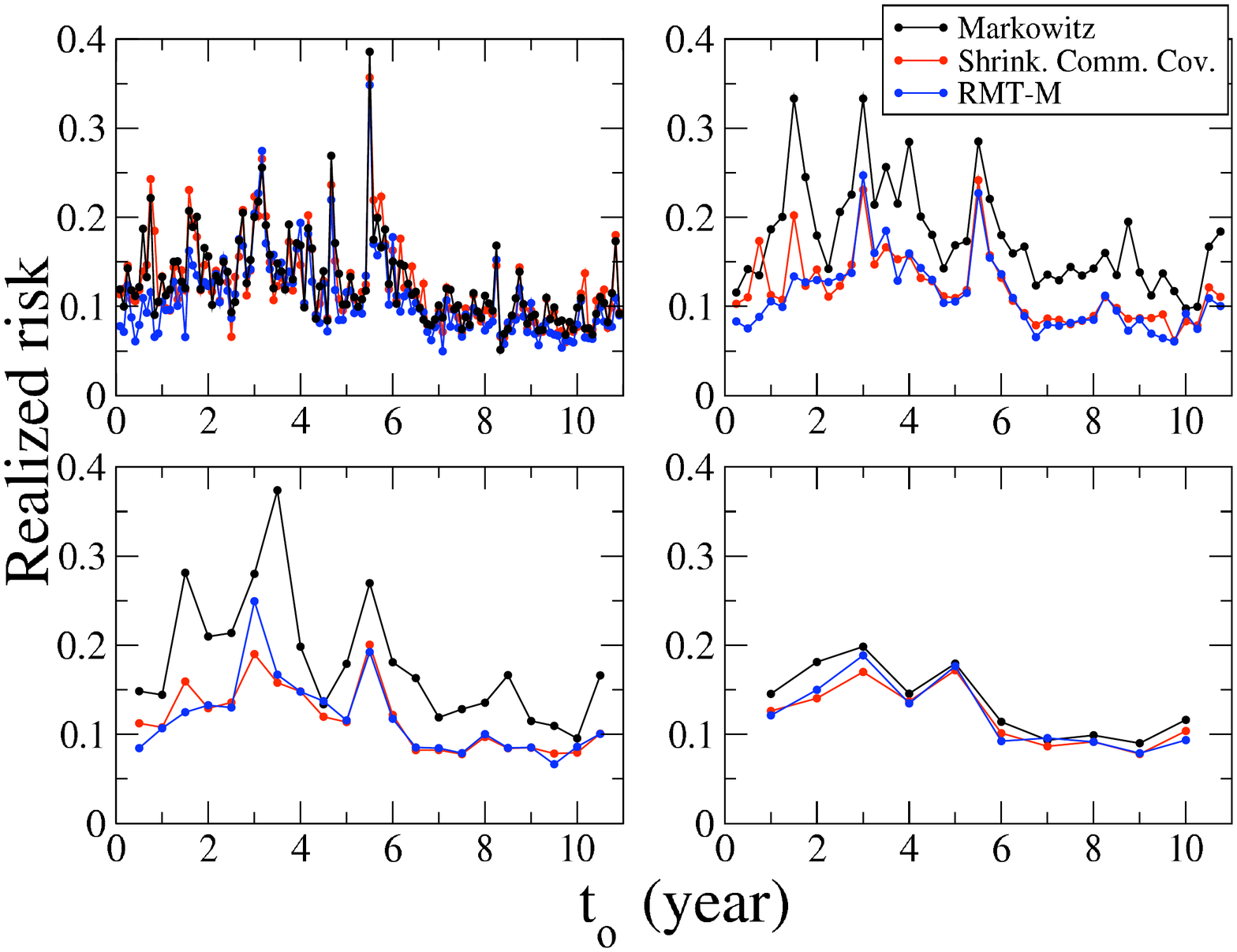}
\end{center}
\caption{Time series of the realized risk $\hat s_p$ over the 11 years of the  Markowitz, the RMT-M, and the shrinkage to common covariance portfolios for a portfolio horizon $T$ equal to $1$ (top left panel), $3$ (top right panel), $6$ (bottom left panel), and $12$ (bottom right panel) months. In these optimizations short selling is allowed.}
\label{timeseries}
\end{figure} 
In Figure \ref{timeseries} we show the  time series of the realized risk as a function of the optimization time $t_0$ for the Markowitz direct optimization and for two representative covariance estimation methods (the shrinkage to common covariance and the RMT-M) when $T=1,3,6$, and $12$ months and short selling is allowed. From the figure it is evident that, for a given method, the temporal fluctuations in the time series of the realized risk are typically larger than the typical differences between the realized risk of the different methods. The same is true if we compare other estimators and also when short selling is not allowed.
The observed high fluctuations in the realized risk indicate that, for a detailed comparison of different portfolio performances, a comparison of the relative differences between portfolio realized risks is more appropriate than a comparison of the average realized risk (averaged over different portfolio selection times). For example, let us consider the yearly case (bottom right panel of Fig. \ref{timeseries}). The realized risk of the Markowitz (black circles) and shrinkage to common covariance (red circles) portfolios averaged over the 11 year time period  are $13.6\%\pm1.3\%$ and $12.1\%\pm1.1\%$, respectively, where errors are standard errors. From these numbers one would conclude that the two methods are equivalent in terms of realized risk. On the contrary, from the time series in the bottom right panel of Fig. \ref{timeseries}, one concludes that the realized risk of the shrinkage to common covariance portfolio is systematically smaller than the one of Markowitz portfolio. In fact, our results show that, for a yearly investment horizon  when short selling is allowed, the shrinkage to common covariance method outperforms all of the other methods.

For these reasons we measure portfolio performances relative to the Markowitz portfolio by means of quantity $1-{\hat{s}_p}/{\hat{s}_p^{(M)}}$ where $\hat{s}_p$ is the realized  risk of the investigated portfolio and $\hat{s}_p^{(M)}$ is the realized risk for the Markowitz portfolio in the same period and conditions. This quantity measures how the investigated portfolio outperforms the Markowitz portfolio (in percentage) in terms of realized risk. 
To assess the statistical robustness of the difference observed between a result obtained with a given covariance estimator and the Markowitz one, we perform a $t$-test to evaluate whether the difference $\hat{s}_p^{(M)} - \hat{s}_p$ has mean value equal to zero.
Similarly, in order to test whether a given portfolio is more reliable than the Markowitz one we perform a $t$-test to evaluate whether
the difference $|\hat{s}_p^{(M)} -s_p^{(M)}|-|\hat{s}_p - s_p|$ is different from zero. Here $s_p$ and $s_p^{(M)}$ are the predicted risk for the investigated and the Markowitz portfolio, respectively.

A quantitative comparison of all the covariance estimator methods is provided in Tables \ref{year}, \ref{sixmonths}, and \ref{onemonth} for the cases $T=1$ year, 6 months, and 1 month, respectively, for both the case when short selling is allowed and when it is not. Since $N=90$, in the first two cases it is  $T>N$, while in the third case it is $T<N$. 

Let us discuss first the case in which short selling is allowed.
Comparing the mean values of $1-\hat{s}_p/\hat{s}_p^ {(M)}$ (third column in the Tables) and the results of the $t$-tests, we conclude that relative portfolio performances depend on the investment horizon $T$. For a yearly horizon, all methods except SI and UPGMA outperform the Markowitz portfolio and the best method is shrinkage to common covariance (as already noted above) which has a realized risk an 11\% smaller on average than the Markowitz portfolio. Note that when $T$ is equal to one year, RMT-M also performs similarly well. In fact the average realized risk for this method is 10.4\% smaller than the Markowitz one. However for lower time horizons a different pattern emerges. When $T=6$ months (Table \ref{sixmonths}), all portfolios perform equally well compared to the Markowitz portfolio, being roughly $33\%$ less risky than the Markowitz portfolio. When $T=1$ month  (see Table \ref{onemonth}), all methods except shrinkage to common covariance outperform Markowitz direct optimization. The spectral methods SI, RMT-0, and RMT-M perform the best and equally well. Among shrinkage methods, shrinkage to SI and shrinkage to constant correlation perform almost as well as the spectral methods, while the shrinkage to common covariance portfolio is the worst, having a realized risk which is statistically indistinguishable from the Markowitz portfolio. 
By considering the reliability which is given in the last column of the Tables, we conclude that 
all the methods outperform Markowitz with a single exception observed for the SI covariance estimator when $T=1$ year. Again the degree of improvement is enhanced when $T=6$ months.

We now consider the no short selling case. As anticipated in the previous discussion,
for $T>N$ all portfolios have similar realized risks and the observed values are quite close  to those observed in the absence of no short selling constraint. This is confirmed by the results shown in the bottom part of Tables \ref{year} and \ref{sixmonths}.  For $T=1$ year the quantity $1-\hat{s}_p/\hat{s}_p^{(M)}$ is essentially consistent with zero for all portfolios. When $T=6$ months only the shrinkage to single index estimator performs slightly better than Markowitz direct optimization at a 5\%  confidence level. For $T=1$ month (Table \ref{onemonth}) a different result emerges. In fact, all portfolios have a significantly smaller realized risk than the Markowitz portfolio. The only notable exception is the shrinkage to common covariance portfolio that presents the same (bad) performance as  the Markowitz portfolio. The best results for the realized risk are observed for hierarchical clustering methods and for the shrinkage to constant correlation method. Moreover the spectral methods perform slightly worse than the others with respect to risk forecasting.

Note that when $T/N\approx1$ the bad performance of Markowitz portfolio, observed when short selling constraints are not imposed, is no longer present. The no short selling constraint makes the Markowitz optimization procedure essentially equivalent to an optimization procedure that has been performed with more robust covariance estimators. Again this observation is in agreement with the conclusion that  imposing no short selling constraint on the portfolio optimization procedure is somehow equivalent to minimize estimation errors in the input to the optimization problem \cite{Jagannathan2003}. 

\begin{table}
\centering
\caption{Different portfolio performance measures that combine (annualized) predicted $s_p$ and realized $\hat{s}_p$ risks. 10 different methods are compared for an horizon of $T=1$ year.  The numbers are average over the different portfolios and the errors are standard errors. 
For $\hat{s}_p$ and $|\hat{s}_p-s_p|$ we report the result of a $t$-test evaluating whether the difference of each quantity with the corresponding quantity for the Markowitz portfolio has mean value equal to zero. The $p$-value of the null hypothesis is below a 1\% threshold when the symbol ** is present while is below 5\% when the symbol * is present.}
\begin{tabular}{|l|rr|r|r|}\hline\hline
Year -- s.s. & $s_p$& $\hat{s}_p$ & $1-\frac{\hat{s}_p}{\hat{s}_p^{(M)}}$ & $|\hat{s}_p-s_p|$ \\ 
\hline\hline
Markowitz   	    & 6.97 $\pm$ 0.63& 13.6 $\pm$ 1.3~~~~& 0 $\pm$ 0& 6.7 $\pm$ 1.1~~~~\\ 
\hline
SI    		    & 5.94 $\pm$ 0.41& 13.2 $\pm$ 1.3~--~& 2.7 $\pm$ 5.0& 7.2 $\pm$ 1.2~--~\\ 
RMT-0		    & 7.18 $\pm$ 0.67& 12.4 $\pm$ 1.2**& 9.5 $\pm$ 2.5& 5.2 $\pm$ 1.1**\\ 
RMT-M		    & 7.24 $\pm$ 0.68& 12.2 $\pm$ 1.2**& 10.4 $\pm$ 2.4& 5.1 $\pm$ 1.0**\\ 
\hline
UPGMA		    & 8.23 $\pm$ 0.88& 13.0 $\pm$ 1.3~--~& 5.0 $\pm$ 2.3& 4.8 $\pm$ 1.1**\\ 
WPGMA		    & 7.88 $\pm$ 0.82& 12.6 $\pm$ 1.3*~~& 7.6 $\pm$ 2.6& 4.8 $\pm$ 1.1**\\ 
Hausdorff	    & 7.57 $\pm$ 0.80& 12.3 $\pm$ 1.2*~~& 9.3 $\pm$ 3.0& 4.75 $\pm$ 0.99**\\ 
\hline
Shr. to SI     & 7.59 $\pm$ 0.70& 12.3 $\pm$ 1.1**& 9.09 $\pm$ 0.90& 4.76 $\pm$ 0.98**\\ 
Shr. C. Cov.  & 10.54 $\pm$ 0.91& 12.1 $\pm$ 1.1**& 11.0 $\pm$ 1.7& 2.57 $\pm$ 0.69**\\ 
Shr. C. Corr.& 8.33 $\pm$ 0.81& 12.8 $\pm$ 1.2**& 6.3 $\pm$ 1.0& 4.5 $\pm$ 1.0**\\ 
 \hline
 \hline
 Year --  no s.s. & $s_p$& $\hat{s}_p$ & $1-\frac{\hat {s}_p}{\hat {s}_p^{(M)}}$& $|\hat{s}_p-s_p|$\\  
\hline\hline
Markowitz   	    & 9.46 $\pm$ 0.88& 12.7 $\pm$ 1.2~~~~& 0 $\pm$ 0& 4.06 $\pm$ 0.93~~~~\\ 
\hline
SI    		    & 7.90 $\pm$ 0.64& 12.9 $\pm$ 1.2~--~& -2.2 $\pm$ 3.0& 5.5 $\pm$ 1.2~--~\\ 
RMT-0		    & 9.18 $\pm$ 0.84& 12.8 $\pm$ 1.2~--~& -0.34 $\pm$ 0.97& 4.33 $\pm$ 0.98~--~\\ 
RMT-M		    & 9.08 $\pm$ 0.83& 12.8 $\pm$ 1.2~--~& 0.07 $\pm$ 0.95& 4.33 $\pm$ 0.98~--~\\ 
\hline
UPGMA		    & 9.9 $\pm$ 1.0& 12.9 $\pm$ 1.3~--~& -0.70 $\pm$ 0.98& 3.93 $\pm$ 0.97~--~\\ 
WPGMA		    & 9.01 $\pm$ 0.89& 12.7 $\pm$ 1.2~--~& 0.2 $\pm$ 1.5& 4.11 $\pm$ 0.98~--~\\ 
Hausdorff	    & 8.68 $\pm$ 0.91& 12.5 $\pm$ 1.1~--~& 1.7 $\pm$ 2.1& 4.14 $\pm$ 0.95~--~\\ 
\hline
Shr. to SI     & 9.35 $\pm$ 0.85& 12.6 $\pm$ 1.1~--~& 0.75 $\pm$ 0.42& 4.01 $\pm$ 0.93~--~\\ 
Shr. C. Cov.  & 11.7 $\pm$ 1.0& 12.2 $\pm$ 1.1~--~& 3.4 $\pm$ 1.9& 2.40 $\pm$ 0.72~--~\\ 
Shr. C. Corr.& 10.05 $\pm$ 0.98& 12.8 $\pm$ 1.2~--~& -0.43 $\pm$ 0.90& 3.92 $\pm$ 0.92~--~\\ 
\hline 
 \hline
  \end{tabular}
\label{year}
\end{table}

\begin{table}
\centering
\caption{Different portfolio performance measures that combine (annualized) predicted $s_p$ and realized $\hat {s}_p$ risks. 10 different methods are compared for an horizon of $T=6$ months.  The numbers are average over the different portfolios and the errors are standard errors. For $\hat{s}_p$ and $|\hat{s}_p-s_p|$ we report the result of a $t$-test evaluating whether the difference of each quantity with the corresponding quantity for the Markowitz portfolio has mean value equal to zero. The $p$-value of the null hypothesis is below a 1\% threshold when the symbol ** is present while is below 5\% when the symbol * is present.}
\begin{tabular}{|l|rr|r|r|}\hline\hline
6 months --  s.s. & $s_p$& $\hat{s}_p$ & $1-\frac{\hat{s}_p}{\hat{s}_p^{(M)}}$ & $|\hat{s}_p-s_p|$\\ 
\hline\hline
Markowitz   	  & 4.23 $\pm$ 0.30& 18.1 $\pm$ 1.5~~~~~~&0 $\pm$ 0~~~~~~& 13.9 $\pm$ 1.4~~~~~~\\
\hline
SI    		   & 5.52 $\pm$ 0.33& 12.05 $\pm$ 0.92**& 31.3 $\pm$ 3.0& 6.53 $\pm$ 0.83**\\ 
RMT-0		    & 6.10 $\pm$ 0.42& 11.91 $\pm$ 0.96**& 32.4 $\pm$ 3.3& 5.81 $\pm$ 0.82**\\ 
RMT-M		    & 6.17 $\pm$ 0.43& 11.80 $\pm$ 0.95**& 33.0 $\pm$ 3.2& 5.63 $\pm$ 0.82**\\
\hline
UPGMA		    & 7.46 $\pm$ 0.57& 12.12 $\pm$ 0.91**& 31.1 $\pm$ 3.1& 4.66 $\pm$ 0.76**\\
WPGMA		    & 7.22 $\pm$ 0.56& 11.86 $\pm$ 0.86**& 32.3 $\pm$ 3.1& 4.65 $\pm$ 0.74**\\
Hausdorff	   & 6.48 $\pm$ 0.55& 11.82 $\pm$ 0.82**& 32.4 $\pm$ 3.0& 5.34 $\pm$ 0.77**\\ 
\hline
Shr. to SI     & 6.41 $\pm$ 0.43& 11.72 $\pm$ 0.82**& 33.4 $\pm$ 2.4& 5.30 $\pm$ 0.65**\\ 
Shr. C. Cov.  & 10.77 $\pm$ 0.76& 11.73 $\pm$ 0.80**& 33.2 $\pm$ 2.4& 2.82 $\pm$ 0.55**\\ 
Shr. C. Corr.& 7.51 $\pm$ 0.53& 12.05 $\pm$ 0.88**& 31.7 $\pm$ 2.7& 4.54 $\pm$ 0.67**\\ 
 \hline
 \hline
6 months --  no s.s. & $s_p$& $\hat{s}_p$ & $1-\frac{\hat {s}_p}{\hat{s}_p^{(M)}}$ & $|\hat{s}_p-s_p|$\\ 
\hline\hline
Markowitz   	   & 8.57 $\pm$ 0.63& 11.85 $\pm$ 0.87& 0 $\pm$ 0& 3.94 $\pm$ 0.69~~~\\
\hline
SI    		    & 7.40 $\pm$ 0.52& 11.98 $\pm$ 0.86~--~& -1.7 $\pm$ 1.5& 4.92 $\pm$ 0.78*~~\\  
RMT-0		    & 8.27 $\pm$ 0.62& 11.83 $\pm$ 0.86~--~& -0.1 $\pm$ 1.0& 4.17 $\pm$ 0.72~--~\\ 
RMT-M		   & 8.20 $\pm$ 0.61& 11.81 $\pm$ 0.86~--~& 0.1 $\pm$ 1.0& 4.21 $\pm$ 0.72~--~\\
\hline
UPGMA		    & 9.19 $\pm$ 0.72& 11.83 $\pm$ 0.89~--~& 0.26 $\pm$ 0.96& 3.57 $\pm$ 0.72~--~\\ 
WPGMA		    & 8.42 $\pm$ 0.67& 11.79 $\pm$ 0.87~--~& 0.4 $\pm$ 1.0& 3.75 $\pm$ 0.78~--~\\
Hausdorff	    & 7.45 $\pm$ 0.67& 12.04 $\pm$ 0.82~--~& -2.5 $\pm$ 1.5& 4.88 $\pm$ 0.83*~~\\  
\hline
Shr. to SI     & 8.48 $\pm$ 0.61& 11.69 $\pm$ 0.87*~~& 1.31 $\pm$ 0.51& 3.87 $\pm$ 0.71~--~\\ 
Shr. C. Cov.  & 11.79 $\pm$ 0.84& 11.84 $\pm$ 0.85~--~& -0.6 $\pm$ 2.2& 3.30 $\pm$ 0.63~--~\\ 
Shr. C. Corr & 9.48 $\pm$ 0.71& 11.86 $\pm$ 0.93~--~& 0.5 $\pm$ 1.1& 3.42 $\pm$ 0.73*~~\\ 
\hline 
 \hline
 \end{tabular}
\label{sixmonths}
\end{table}

\begin{table}
\centering
\caption{Different portfolio performance measures that combine predicted $s_p$ and the realized $\hat {s}_p$ annualized risk. 10 different methods are compared for an horizon of $T=1$ month.  The numbers are average over the different portfolios and the errors are standard errors. For $\hat{s}_p$ and $|\hat{s}_p-s_p|$ we report the result of a $t$-test evaluating whether the difference of each quantity with the corresponding quantity for the Markowitz portfolio has mean value equal to zero. The $p$-value of the null hypothesis is below a 1\% threshold when the symbol ** is present while is below 5\% when the symbol * is present.}
\begin{tabular}{|l|rr|r|r|}\hline\hline
 Month --  s.s. & $s_p$& $\hat{s}_p$ & $1-\frac{\hat{s}_p}{\hat{s}_p^{(M)}}$ & $|\hat{s}_p-s_p|$\\ 
\hline\hline
Markowitz   	    & 0 $\pm$ 0& 12.59 $\pm$ 0.41~~~~& 0 $\pm$ 0~~~& 12.59 $\pm$ 0.41~~~\\ 
\hline
SI    		    & 4.15 $\pm$ 0.12& 11.00 $\pm$ 0.42**& 12.1 $\pm$ 1.5& 6.85 $\pm$ 0.37**\\ 
RMT-0		    & 3.84 $\pm$ 0.11& 10.94 $\pm$ 0.39**& 12.5 $\pm$ 1.4& 7.10 $\pm$ 0.34**\\ 
RMT-M		    & 3.90 $\pm$ 0.12& 10.91 $\pm$ 0.39**& 12.8 $\pm$ 1.4& 7.01 $\pm$ 0.34**\\ 
\hline
UPGMA		    & 5.01 $\pm$ 0.17& 11.66 $\pm$ 0.45**& 6.6 $\pm$ 2.1& 6.65 $\pm$ 0.38**\\ 
WPGMA		    & 4.74 $\pm$ 0.17& 11.44 $\pm$ 0.44**& 8.3 $\pm$ 1.9& 6.70 $\pm$ 0.37**\\ 
Hausdorff	    & 4.98 $\pm$ 0.17& 11.62 $\pm$ 0.45**& 7.0 $\pm$ 2.1& 6.64 $\pm$ 0.37**\\ 
\hline
Shr. to SI     & 3.48 $\pm$ 0.15& 11.04 $\pm$ 0.39**& 11.8 $\pm$ 1.2& 7.57 $\pm$ 0.35**\\ 
Shr. C. Cov.  & 13.1 $\pm$ 0.47& 12.44 $\pm$ 0.42~--~& 0.5 $\pm$ 1.5& 3.64 $\pm$ 0.30**\\ 
Shr. C. Corr.& 5.87 $\pm$ 0.20& 11.56 $\pm$ 0.45**& 7.4 $\pm$ 1.9& 5.70 $\pm$ 0.37**\\ 
\hline
\hline 
Month-   no s.s. & $s_p$& $\hat {s}_p$ & $1-\frac{\hat {s}_p}{\hat {s}_p^{(M)}}$ & $|\hat{s}_p-s_p|$\\ 
\hline\hline
Markowitz   	    & 4.38 $\pm$ 0.24& 13.09 $\pm$ 0.52~~~& 0 $\pm$ 0~~~& 8.73 $\pm$ 0.53~~~~\\ 
\hline
SI    		    & 5.60 $\pm$ 0.20& 11.60 $\pm$ 0.44**& 9.3 $\pm$ 1.4& 6.04 $\pm$ 0.39**\\ 
RMT-0		    & 5.48 $\pm$ 0.21& 11.57 $\pm$ 0.42**& 9.5 $\pm$ 1.2& 6.11 $\pm$ 0.38**\\ 
RMT-M		    & 5.49 $\pm$ 0.21& 11.54 $\pm$ 0.42**& 9.7 $\pm$ 1.2& 6.07 $\pm$ 0.38**\\ 
\hline
UPGMA		    & 7.11 $\pm$ 0.25& 11.45 $\pm$ 0.44**& 10.8 $\pm$ 1.3& 4.54 $\pm$ 0.37**\\ 
WPGMA		    & 6.15 $\pm$ 0.22& 11.48 $\pm$ 0.44**& 10.6 $\pm$ 1.2& 5.39 $\pm$ 0.38**\\ 
Hausdorff	    & 6.73 $\pm$ 0.23& 11.53 $\pm$ 0.43**& 10.3 $\pm$ 1.2& 4.87 $\pm$ 0.34**\\ 
\hline
Shr. to SI     & 5.72 $\pm$ 0.21& 11.76 $\pm$ 0.43**& 8.64 $\pm$ 0.91& 6.06 $\pm$ 0.38**\\ 
Shr. C. Cov.  & 13.39 $\pm$ 0.48& 12.74 $\pm$ 0.44~--~& -2.6 $\pm$ 2.6& 3.76 $\pm$ 0.30**\\ 
Shr. C. Corr.& 8.20 $\pm$ 0.29& 11.56 $\pm$ 0.47**& 10.3 $\pm$ 1.4& 3.93 $\pm$ 0.35**\\ 
\hline
\hline
\end{tabular}
\label{onemonth}
\end{table}

\subsection{Portfolio diversification}
\begin{table}
\centering
\caption{Absolute and relative participation ratio measure $N_{eff}$ of the portfolios obtained with the 10 covariance estimators for different horizons of $T=1$, 6 and 12 months. Short selling is not allowed. The numbers are average over the different portfolios and the errors are standard errors. For $N_{eff}$ we report the result of a $t$-test evaluating whether the difference with the corresponding quantity for the Markowitz portfolio has mean value equal to zero.The $p$-value of the null hypothesis is below a 1\% threshold when the symbol ** is present while is below 5\% when the symbol * is present.}
\begin{tabular}{l|rr|rr|rr}
\hline
   & One  & month~~~~~~ & Six & months~~~~~~ & One & year~~~~~~\\
  &$N_{eff}$ & $\frac{N_{eff}}{N_{eff}^{(M)}}-1$&$N_{eff}$ & $\frac{N_{eff}}{N_{eff}^{(M)}}-1$&$N_{eff}$ & $\frac{N_{eff}}{N_{eff}^{(M)}}-1$\\
\hline\hline
Markowitz   	    & 6.80 $\pm$ 0.22~~~   &   0.0$\pm$ 0.0 & 9.8 $\pm$ 1.0~~~&    0.0$\pm$0.0 & 9.9 $\pm$ 1.5~~~   & 0.0$\pm$ 0.0~~~	   \\ 
\hline
SI    		             & 14.91 $\pm$ 0.98** & 104.0 $\pm$ 8.4  & 14.0 $\pm$ 2.1**   & 36.8 $\pm$ 7.5 & 13.8 $\pm$ 2.7  & 33.4 $\pm$ 9.2*~~  \\ 
RMT-0		    & 13.45 $\pm$ 0.80**  & 85.4 $\pm$ 6.2 & 11.2 $\pm$ 1.3** & 13.4 $\pm$ 2.7 & 10.6 $\pm$ 1.7  & 6.8 $\pm$ 4.0~--~     \\ 
RMT-M		    & 13.63 $\pm$ 0.81** & 87.9 $\pm$ 6.2 & 11.6 $\pm$ 1.3** & 16.9 $\pm$ 2.9 & 10.9 $\pm$ 1.7  & 10.1 $\pm$ 4.0~--~    \\ 
\hline
UPGMA		    & 8.90 $\pm$ 0.44**   & 26.5 $\pm$ 3.5 & 10.2 $\pm$ 1.1** & 5.1 $\pm$ 3.7  & 10.7 $\pm$ 1.8  & 6.7 $\pm$ 4.6~--~  \\ 
WPGMA		    & 11.62 $\pm$ 0.53** & 67.6 $\pm$ 4.3 & 12.1 $\pm$ 1.1** & 26.3 $\pm$ 5.2 & 13.0 $\pm$ 1.9    & 30.5 $\pm$ 3.6**  \\ 
Hausdorff	             & 9.55 $\pm$ 0.34**  & 42.4 $\pm$ 3.3 & 13.1 $\pm$ 1.4** & 36.0 $\pm$ 5.5   & 13.0 $\pm$ 1.8    & 34.9 $\pm$ 4.6**   \\ 
\hline
Shr. to SI     & 11.7 $\pm$ 0.67**  & 60.9 $\pm$ 5.1 & 11.3 $\pm$ 1.4** & 11.8 $\pm$ 2.2 & 10.7 $\pm$ 1.8  & 7.3 $\pm$ 1.8**    \\ 
Shr. C. Cov.  & 37.3 $\pm$ 1.4**   & 530 $\pm$ 45 	& 18.9 $\pm$ 1.5** & 159 $\pm$ 64   & 15.5 $\pm$ 1.8  & 100 $\pm$ 51**    \\ 
Shr. C. Corr.& 7.64 $\pm$ 0.43**  & 7.5 $\pm$ 3.8  & 10.1 $\pm$ 1.2~--~ & -0.1 $\pm$ 2.6 & 10.0 $\pm$ 1.7    & -1.3 $\pm$ 2.8~--~   \\ 
\hline
\hline
\end{tabular}
\label{neffnss}
\end{table}
\begin{table}
\centering\caption{Absolute and relative participation ratio measure $N_{90}$ of the portfolios obtained with the 10 covariance estimators for different horizons of $T=1$, 6 and 12 months. Short selling is not allowed. The numbers are average over the different portfolios and the errors are standard errors. For $N_{90}$ we report the result of a $t$-test evaluating whether the difference with the corresponding quantity for the Markowitz portfolio has mean value equal to zero. The $p$-value of the null hypothesis is below a 1\% threshold when the symbol ** is present while is below 5\% when the symbol * is present.}
\begin{tabular}{l|rr|rr|rr}
\hline
Short selling   & One & month& Six & months & One & year\\
  &$N_{90}$ & $\frac{N_{90}}{N_{90}^{(M)}}-1$&$N_{90}$ & $\frac{N_{90}}{N_{90}^{(M)}}-1$&$N_{90}$ & $\frac{N_{90}}{N_{90}^{(M)}}-1$\\
\hline\hline
Markowitz   	     & 59.41 $\pm$ 0.18~~~  & 0.0 $\pm$ 0.0~& 56.81 $\pm$ 0.52~~~ & 0.0 $\pm$ 0.0~& 55.3 $\pm$ 0.99~~~ & 0.0 $\pm$ 0.0~   \\
SI    		     & 52.85 $\pm$ 0.31**  & -10.95 $\pm$ 0.59 & 55.48 $\pm$ 0.71~-- & -2.2 $\pm$ 1.4 & 55.1 $\pm$ 1.2~--~  & -0.3 $\pm$ 1.6  \\
RMT-0		     & 53.87 $\pm$ 0.29**  & -9.23 $\pm$ 0.54  & 55.57 $\pm$ 0.67~-- & -2.1 $\pm$ 1.2 & 55.1 $\pm$ 0.95~--~ & -0.2 $\pm$ 1.6  \\
RMT-M		     & 53.85 $\pm$ 0.29**  & -9.26 $\pm$ 0.54  & 55.38 $\pm$ 0.68*~ & -2.4 $\pm$ 1.2 & 55.1 $\pm$ 0.97~--~ & -0.2 $\pm$ 1.6  \\
UPGMA		     & 52.27 $\pm$ 0.29**  & -11.91 $\pm$ 0.55 & 54.57 $\pm$ 0.49** & -3.8 $\pm$ 1.1 & 55.6 $\pm$ 0.97~--~ & 0.7 $\pm$ 1.9   \\
WPGMA		     & 51.64 $\pm$ 0.28**  & -12.96 $\pm$ 0.56 & 54.14 $\pm$ 0.67** & -4.6 $\pm$ 1.3 & 54.9 $\pm$ 1.0~--~    & -0.6 $\pm$ 2.0     \\
Hausdorff	     & 52.03 $\pm$ 0.26**  & -12.31 $\pm$ 0.52 & 52.48 $\pm$ 0.70**  & -7.6 $\pm$ 1.2 & 53.7 $\pm$ 1.1~--~  & -2.7 $\pm$ 2.1   \\
Shr. to SI      & 53.45 $\pm$ 0.29**  & -9.97 $\pm$ 0.50   & 54.38 $\pm$ 0.63** & -4.2 $\pm$ 1.0   & 55.0 $\pm$ 1.1~--~    & -0.5 $\pm$ 1.5  \\
Shr. C. Cov.   & 60.89 $\pm$ 0.35**  & 2.57 $\pm$ 0.61   & 57.81 $\pm$ 0.49~-- & 1.9 $\pm$ 1.3  & 57.2 $\pm$ 1.0~--~    & 3.6 $\pm$ 2.1    \\
Shr. C. Corr  & 52.97 $\pm$ 0.31**  & -10.71 $\pm$ 0.62 & 53.95 $\pm$ 0.64** & -5.0 $\pm$ 1.1   & 54.6 $\pm$ 1.0~--~    & -1.24 $\pm$ 0.94  \\
\hline
\hline
No short selling   & One & month& Six & months & One & year\\
  &$N_{90}$ & $\frac{N_{90}}{N_{90}^{(M)}}-1$&$N_{90}$ & $\frac{N_{90}}{N_{90}^{(M)}}-1$&$N_{90}$ & $\frac{N_{90}}{N_{90}^{(M)}}-1$\\
\hline\hline
Markowitz   	     & 8.40 $\pm$ 0.19~~ 	 & 0.0 $\pm$ 0.0~   & 12.81 $\pm$ 1.00~~   & 0.0 $\pm$ 0.0~   & 13.4 $\pm$ 1.5~~  & 0.0 $\pm$ 0.0~    \\
\hline
SI    		     & 18.9 $\pm$ 1.1** 	 & 113.3 $\pm$ 8.2 & 17.0 $\pm$ 2.0**      & 31.1 $\pm$ 7.6  & 16.4 $\pm$ 2.6~--  & 18.7 $\pm$ 8.0     \\
RMT-0		     & 17.21 $\pm$ 0.85**  & 95.9 $\pm$ 6.1  & 13.8 $\pm$ 1.2*~  & 8.9 $\pm$ 4.0     & 13.4 $\pm$ 1.7~--  & -0.7 $\pm$ 3.7   \\
RMT-M		     & 17.40 $\pm$ 0.85**   & 98.3 $\pm$ 6.0   & 14.3 $\pm$ 1.2**  & 12.5 $\pm$ 3.9  & 13.9 $\pm$ 1.7~--  & 3.1 $\pm$ 3.3    \\
\hline
UPGMA		     & 11.55 $\pm$ 0.48**  & 33.3 $\pm$ 3.4  & 12.9 $\pm$ 1.1~--  & -0.8 $\pm$ 3.6  & 13.2 $\pm$ 1.9~--  & -4.5 $\pm$ 5.0     \\
WPGMA		     & 15.39 $\pm$ 0.59**  & 79.8 $\pm$ 4.4  & 15.6 $\pm$ 1.2**  & 23.5 $\pm$ 5.7  & 16.1 $\pm$ 1.7**  & 20.5 $\pm$ 3.4    \\
Hausdorff	              & 12.61 $\pm$ 0.34**  & 51.5 $\pm$ 2.9  & 17.4 $\pm$ 1.4**  & 37.4 $\pm$ 4.9  & 16.4 $\pm$ 1.4**  & 25.7 $\pm$ 4.9    \\
\hline
Shr. to SI      & 15.24 $\pm$ 0.74**  & 72.4 $\pm$ 5.2  & 14.6 $\pm$ 1.4**  & 12.5 $\pm$ 3.0    & 14.4 $\pm$ 1.9~--  & 5.7 $\pm$ 2.7    \\
Shr. C. Cov.   & 37.4 $\pm$ 1.2** 	 & 363 $\pm$ 20    & 21.3 $\pm$ 1.3**  & 85 $\pm$ 22     & 18.8 $\pm$ 1.7**  & 46 $\pm$ 10       \\
Shr. C. Corr  & 10.00 $\pm$ 0.51** 	 & 14.3 $\pm$ 3.9  & 12.7 $\pm$ 1.3~--  & -4.2 $\pm$ 4.0   & 13.5 $\pm$ 1.9~--  & -1.8 $\pm$ 4.8     \\
\hline
\hline
\end{tabular}
\label{n90}
\end{table}

One further aspect to investigate concerns the degree of diversification of portfolios. As for the realized risk, for the Markowitz direct optimization and for any given covariance estimator, we observe large fluctuations of the participation ratio as the portfolio estimation time $t_0$ varies. We therefore consider both the mean and the standard error of $N_{eff}$ for each method across time and the mean value of ${N_{eff}}/{N_{eff}^{(M)}}-1$ in percentage, where $N_{eff}^{(M)}$ is the participation ratio for the Markowitz portfolio. This variable is a relative measure that quantifies the portfolio diversification with respect to the diversification of the benchmark Markowitz portfolio.
Also in this case we perform a $t$-test in order to evaluate whether the observed difference  $N^{(M)}_{eff} - N_{eff}$ is compatible with a null hypothesis assuming that its mean value is zero. 

In Table \ref{neffnss} we report the average and standard error for $N_{eff}$ and ${N_{eff}}/{N_{eff}^{(M)}}-1$ for the 10 optimization methods and for $T=1$ month, 6 months, and 1 year, together with the related results for the $t$-test. The Table shows a different behavior at different values of the investment time window $T$. Specifically, at $T=1$ month all methods present a participation ratio which is higher than the one observed for Markowitz direct optimization. When $T=6$ months all methods still outperform Markowitz with the exception of the shrinkage to constant correlation. When $T=1$ year there are still several methods that outperforms Markowitz, namely SI, WPGMA, Hausdorff, shrinkage to single index and shrinkage to common covariance. 
The method with the highest participation ratio at any time horizon is the shrinkage to common covariance. For example, when $T=1$ month it has a participation ratio which is $530\%$ higher than the Markowitz portfolio on average. This high diversification is not shared with the other two shrinkage methods. This is probably due to the fact that the target matrix of the shrinkage to common covariance assumes that all the stocks are equivalent. SI among the spectral methods and WPGMA among the  hierarchical clustering methods have the highest participation ratio of the other classes of covariance estimators.

In the above discussion, we have used $N_{eff}$ to quantify the portfolio diversification under no short selling constraint. In fact, we have already discussed that this indicator is not meaningful when short selling is allowed. For this reason, we now consider the second participation ratio indicator, $N_{90}$, introduced above. Table \ref{n90} reports the mean and the standard error of $N_{90}$ for each method averaged across investment time and, as before, a relative measure both when short selling is allowed and when it is forbidden. 
We also perform a $t$-test to evaluate whether the difference $N^{(M)}_{90}-N_{90}$ has a mean value significantly different from zero.

When short selling is not allowed $N_{90}$ gives results very close to those observed for $N_{eff}$. In fact when $T=1$ month  all the methods give a portfolio more diversified than Markowitz direct optimization. When $T=6$ months all the methods  outperform Markowitz with the exception of shrinkage to constant correlation and UPGMA, whereas when $T=1$ year, only WPGMA, Hausdorff and shrinkage to common covariance still outperform Markowitz. When short selling is allowed, Markowitz direct optimization provides portfolios characterized by a  $N_{90}$ value slightly higher or statistically compatible with the value observed for the other methods. The only exception is shrinkage to common covariance when $T=1$ month but also in this case the difference observed, although statistically validated, is a very small.

In summary, when short selling is allowed the weights have a similar structure independently of the method, and the wealth (positive or negative) is roughly concentrated in $55$ stocks. 
When short selling is not allowed, a large variety of behaviors is observed depending on the method and on the investment time horizon. In general, the shrinkage to common covariance method has the largest participation ratio.

\begin{figure}[ptb]
\begin{center}
\includegraphics[scale=0.5]{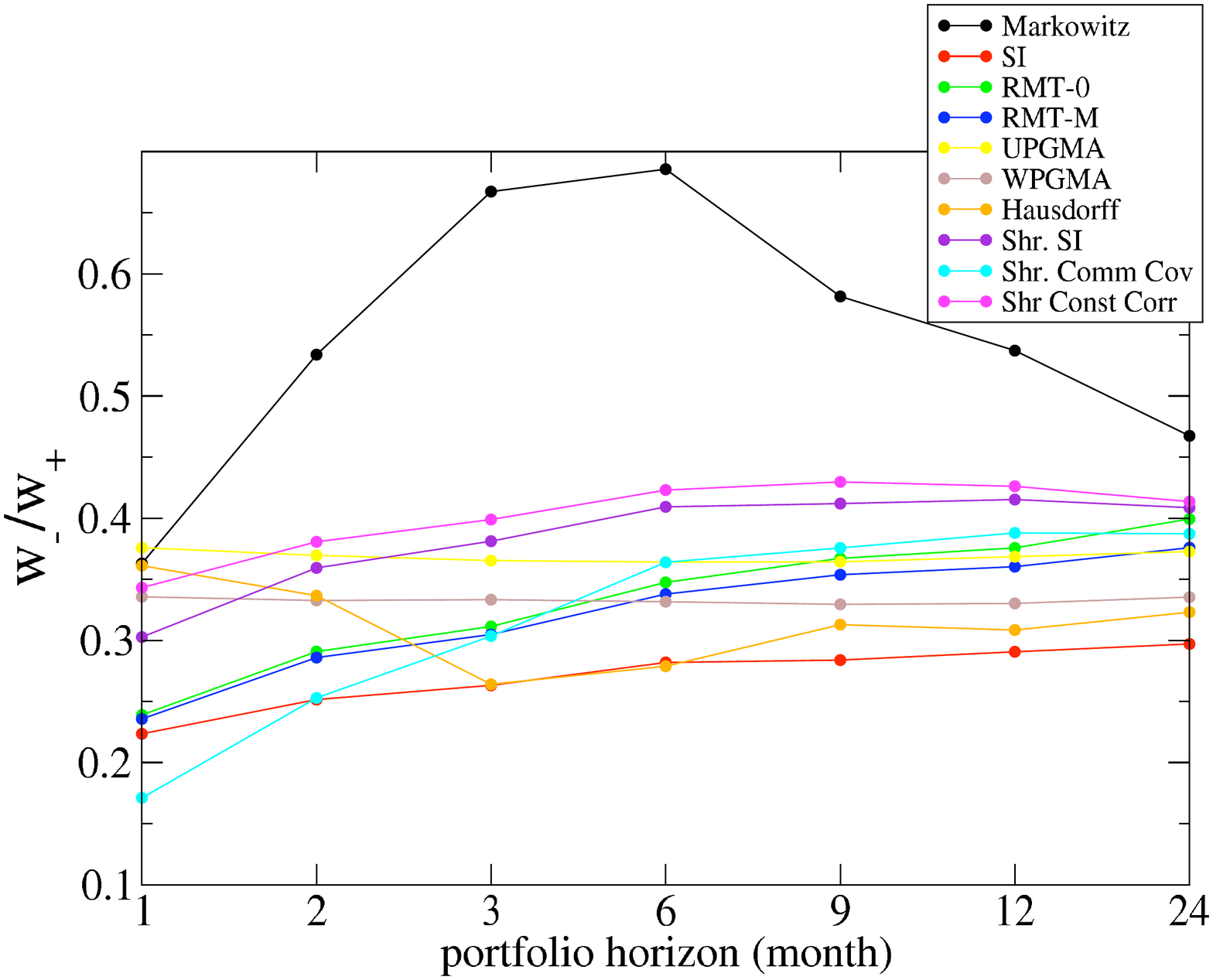}
\end{center}
\caption{Mean value of the ratio $w_{-}/w_{+}$ between the sum of absolute value of negative weights and the sum of positive weights for the portfolios obtained with the 10 different methods as a function of the horizon $T$.}
\label{pesiposinega}
\end{figure}

When short selling is allowed, it is also worth analysing the amount of short selling required by the optimization procedures of the global minimum variance portfolio. To quantify this aspect in Fig. \ref{pesiposinega} we show, for each method, the average value of the ratio $w_{-}/w_{+}$ where $w_{-}$ is the sum of the absolute value of all negative weights present in the portfolio and  $w_{+}$ is the sum of all positive weights.  The ratio $w_{-}/w_{+}$ ranges from 0 (absence of short selling) to about 1 (negative weights of the same size as positive weights).

Fig.  \ref{pesiposinega} shows that Markowitz direct optimization requires the highest fraction of short selling positions. This property is maximal when $T/N\approx1$. All the other methods present a significant lower mean value of $w_{-}/w_{+}$. The specific values depend on the specific covariance estimation method and are slightly affected by the value of the investment horizon $T$. In fact, a slight increase of  $w_{-}/w_{+}$ is observed when $T$ is increasing. The lowest value $ w_{-}/w_{+} \approx 0.28$ is observed for the SI model whereas the highest value $ w_{-}/w_{+} \approx 0.40$ is observed for the shrinkage to constant correlation method. The region of worst performance of the Markowitz direct optimization procedure is therefore associated with the maximal amount of portfolio wealth allocated in stocks that need to be sold short.

These results provide empirical support to the conclusion that Markowitz direct optimization in the presence of short selling suffers of an over exposure to short selling. This over exposure is maximal when $T/N\approx1$ and is progressively mitigated both when $T>N$ and when $T<N$. 
 On the contrary, reducing the estimation errors on the covariance matrix estimation implicitly limits the amount of short selling positions requested in the optimal portfolio. According to the results obtained in Ref. \cite{Jagannathan2003} and to the empirical results obtained in this study, we observe that the reverse is also true. In fact imposing no short selling conditions to the Markowitz optimization reduces the estimation errors in the covariance matrix for any value of $T$, and especially when $T/N\approx1$. 

\section{Conclusions}
The portfolio optimization problem is significantly affected by estimation errors of the covariance matrix. For this reason many estimators alternative to the sample covariance matrix have been proposed in the literature. In this respect, two important and related questions are: (i) which aspects of the portfolio optimization can be improved with improved covariance matrix estimators?
(ii) when, i.e. under which conditions, are improved covariance estimators really useful in enhancing the performance of the corresponding optimal portfolios?
We have investigated these questions by considering $9$ different methods for estimating the covariance matrix and we have quantitatively compared the relative efficiency of the corresponding portfolios with respect to the benchmark Markowitz portfolio on a series of repeated investment exercises over 11 years.  The portfolio optimization has been performed under different  conditions: different estimation-investment horizons $T$, i.e., different values of $T/N$ ($N=90$), and the presence/absence of short selling constraints. Despite the  realized risk and  the degree of portfolio diversification of the resulting portfolios constructed with the different covariance estimators show large fluctuations, relative performances of different methods turn out to be quite persistent over time. Under different market conditions  some persistent behaviors can be observed. For a specific choice of both the length of the estimation-investment horizon and the presence/absence of constraints on sort selling an estimator might be useful in improving a specific aspect of the optimization, but under a different choice the same method might not lead to a significant improvement on the same aspect.

Specifically, when $T/N>1$ various covariance estimators lead to optimal portfolios with similar realized risk and portfolio diversification. In this regime, Markowitz direct optimization has an overall good performance both with and without short selling constraints. While when short selling is allowed a portfolio less risky than the Markowitz one can be obtained by using improved covariance estimators, when short selling is forbidden the investigated estimators are not able to decrease the risk of the portfolio with respect to the Markowitz one. In this last case some covariance estimators lead to higher portfolio diversification. 

On the other hand, when $T/N$ is close to 1, portfolio performances  are greatly influenced by the addition of no short selling constraints. 
Specifically, when short selling is allowed, we observe how the Markowitz direct optimization process has the worst performance. This result is consistent with the theoretical observations given in Ref.  \cite{Jagannathan2003} and with the observation of the divergence of estimation errors of covariance matrix associated with this regime \cite{Pafka2002,Pafka2003,Pafka2005,Kondor2007}. Under this condition  all the investigated covariance estimators provide portfolios with lower realized risk, higher reliability and smaller exposure to short selling. Their performances are quite similar with respect to realized risk, reliability and portfolio diversification but differences are observed with respect to the degree of exposure to short selling. When no short selling constraints are applied, we observe a different scenario. All  covariance estimators lead to portfolios with realized risks and reliabilities that are statistically consistent with those obtained by Markowitz direct optimization. However, portfolios constructed with the investigated methods have a higher degree of diversification  than those observed for the Markowitz direct optimization. This result is consistent with the theoretical and empirical conclusions reached in Ref.  \cite{Jagannathan2003} where it was shown that adding short selling constraints to the Markowitz portfolios  can have the same effect as using a better estimate of the covariance matrix  (using the shrinkage estimator in their case). Our results suggest that indeed this conclusion successfully applies also to other covariance estimators such as the methods investigated in this paper.

When $T/N$ smaller than one, the worst performance with respect to realized risk is obtained for Markowitz direct optimization and shrinkage to common covariance. This result indicates that one should not use the sample covariance matrix in this regime (neither with nor without short selling). Also the use of pseudoinverse gives portfolios with very poor performance. All the other methods lead to portfolios with better performances with respect to realized risk and reliability in realized risk forecasts both in the presence and in the absence of short selling. When the no short selling constraint is imposed, portfolio diversification is better achieved when filtered covariance estimators are used. 
This last observation is also true for the shrinkage to common covariance estimator both when short selling is allowed and when it is forbidden. Indeed this method presents the highest degree of portfolio diversification. It is therefore worth noting that the observation that Markowitz and shrinkage to common covariance portfolios are characterized by similar values of the realized risk does not imply that they have a similar composition. In fact the portfolio obtained with the shrinkage to common covariance method is systematically more diversified. The conclusion reached in Ref. \cite{Jagannathan2003} and empirically observed by us when $T/N\approx 1$ does not seem to hold when $T/N$ is less than one. In fact portfolios obtained with Markowitz direct optimization are characterized by realized risks, reliability of risk forecasts and portfolio diversification that are worse than most of other methods based on covariance estimators also when short selling is forbidden. 

In summary the use of efficient covariance estimators improves different aspects of the portfolio optimization process. The degree of improvement depends on the selected method, the value of the  parameter $T/N$, and the presence or absence of no short selling constraint. The improvements achieved refer to one or more of the following key portfolio indicators: (i) realized risk, (ii) reliability of realized risk predictions, (iii) degree of portfolio diversification and (iv) fraction of short selling when short selling is allowed. 

\section*{Acknowledgments} Authors acknowledge financial support from the PRIN project 2007TKLTSR  ``Indagine di fatti stilizzati e delle strategie risultanti di agenti e istituzioni osservate in mercati finanziari reali ed artificiali".

\end{document}